\documentclass[prb,twocolumn,amsmath,amssymb,floatfix,footinbib,bibnotes]{revtex4}

\usepackage[colorlinks=true,citecolor=blue,linkcolor=magenta]{hyperref}
\usepackage{amsmath}
\usepackage{graphicx}
\usepackage{dsfont}
\usepackage{changepage}
\usepackage{appendix}
\usepackage[T1]{fontenc}
\usepackage[latin1]{inputenc}

\newcommand{\be}{\begin{equation}}
\newcommand{\ee}{\end{equation}}

\def \be{\begin{equation}}
\def \ee{\end{equation}}
\def \ba{\begin{array}}
\def \ea{\end{array}}
\def \bea{\begin{eqnarray}}
\def \eea{\end{eqnarray}}
\def \nn{\nonumber}
\def \half{{1\over 2}}

\def \bq{{\bf q}}
\def \bk{{\bf k}}

\def \W{{\Omega}}
\def \e{{\epsilon}}

\def \L{{\Lambda}}

\def \D{{\Delta}}
\def \d{{\delta}}
\def \w{{\omega}}

\def \e{{\epsilon}}

\def \nd{{^{\vphantom{\dagger}}}}
\def \yd{^\dagger}
\def \av#1{{\langle#1\rangle}}
\def \ket#1{{\,|\,#1\,\rangle\,}}

\def \beas{\begin{eqnarray*}}
\def \eeas{\end{eqnarray*}}

\def \half{{\frac{1}{2}}}

\newcounter{indice}

\begin{document}
\title{Mott criticality and pseudogap in Bose-Fermi mixtures}

\author{Ehud Altman, Eugene Demler, Achim Rosch \\
{$^1$\small \em Department of Condensed Matter Physics, Weizmann Institute of Science, Rehovot 76100, Israel}\\
{$^2$\small \em Department of Physics, Harvard University, Cambridge MA 02138}\\
{$^3$\small \em Institute for Theoretical Physics, University of Cologne, 50937 Cologne, Germany}}

\begin{abstract}
We study the Mott transition of a mixed Bose-Fermi system of ultracold
atoms in an
optical lattice, where the number of (spinless) fermions and bosons adds
up to one atom per lattice, $n_F+n_B=1$. For weak interactions, a
Fermi surface coexists with a Bose-Einstein condensate while for
strong interaction the system is incompressible but still
characterized by a Fermi surface of composite fermions.
At the critical point, the spectral function of the fermions, $A(\bk,\w)$, exhibits a pseudo-gapped behavior, rising as $|\omega|$ at the Fermi momentum, while in the Mott phase it is fully gapped.
Taking into account the interaction between the critical modes leads at very low temperatures either to  p-wave pairing or the transition is driven weakly first order.
The same mechanism should also be important in antiferromagnetic metals with a small Fermi surface.
\end{abstract}
\maketitle
A recent experiment with an ultra-cold mixture of bosonic and
fermionic Yb atoms in an optical lattice \cite{Sugawa2011} has found a
remarkable quantum phase that can be described as a mixed Mott
insulator. Such a state\cite{Lewenstein2004,Massignan2010,Eckardt2010a} is established in the strongly interacting
regime when the average site occupation of the bosons and fermions
together is an integer, $n_B+n_F =0,1,...$.  While the state is
incompressible and hence fluctuations of the total density are gapped,
the fermions can still
move around by exchanging with the spinless bosons. Hence the mobile
objects are bound states of a fermionic atom and a bosonic
hole. Depending on their effective interactions,
these bound states can form a number of different phases,
including a Fermi liquid or a paired condensate. But those
are rather strange fluids, made of composite fermions that carry zero
net particle number. Accordingly, the spectral function of the
original fermionic atoms will not display a quasi-particle peak. This
phase, established for sufficiently strong interactions, should be
contrasted with the weakly interacting limit where the fermionic atoms
form a conventional Fermi sea coexisting with a Bose condensate(BEC) of the
other species.
In this paper, we investigate the quantum phase transition from the incompressibe mixed Mott state to the compressible metal/BEC phase  and the fate of the Fermi surface across the transition.

In most solid state systems the Mott quantum critical point from  a metal to an
insulating state is masked by
antiferromagnetism. In cases where frustration suppresses magnetism,
however, it has been argued that a direct transition from a metallic
phase to an insulating and incompressible U(1) spin liquid is possible
\cite{SSLee,Senthil2008,Senthil2008a,Podolsky}. Yet the understanding of this transition remains rudimentary and is unconfirmed by experiment.
We argue that with ultracold mixtures of bosons and fermions it is possible to study a similar
transition directly. An
essential element of the transition to the U(1) spin liquid, the
coupling to a deconfined $U(1)$ gauge field, is, however, missing.

\begin{figure}[t]
\centerline{\resizebox{0.7\linewidth}{!}{\includegraphics{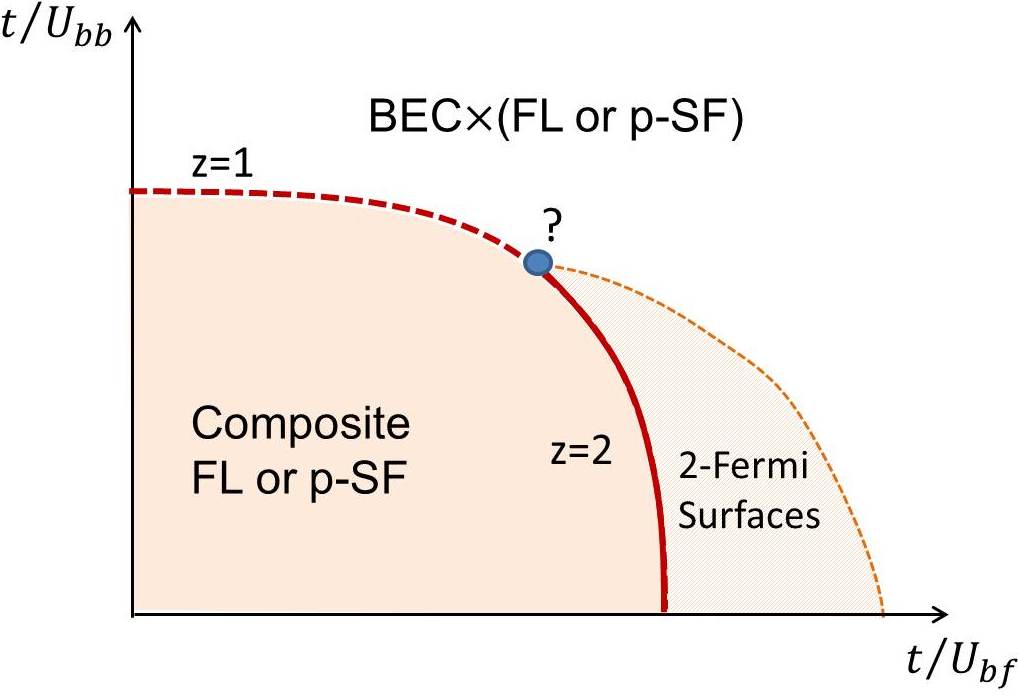}}}
\caption{Schematic phase diagram of the Bose Fermi mixture at combined integer filling. At weak interaction the bosons form a BEC and the fermions form a Fermi liquid, unstable at very low temperature to p-wave pairing. At strong interactions the system goes into the mixed Mott phase, in which composite neutral fermions (with respect to total density) still exist as low energy degrees of freedom. Depending on parameters they can form either a Fermi liquid (FL) a p-wave superfluid or phase separate. The nature of the Mott critical point depends on how it is approached, by tuning the boson-fermion or the boson-boson interactions.}
\label{fig:schem}
\end{figure}

{\em Model --} For simplicity we confine ourselves to  spinless fermions mixed with a single species of bosons in three dimensions ($d=3$), described by the generalized Hubbard model
\bea
H&=&-t_b\sum_\av{ij} (b\yd_i b\nd_j +\text{H.c.})-t_f\sum_\av{ig} (f\yd_i f\nd_j + \text{H.c.})\nn\\
&&+ \half U_{bb}\sum_i n_{bi} (n_{bi}-1) + U_{bf}\sum_i n_{bi} n_{fi}.
\label{Hmic}
\eea Here $b\yd_i$ and $c\yd_i$ create bosonic and fermionic atoms
respectively on site $i$ of the lattice and $n_{bi}$ ($n_{fi}$) are
the bosonic (fermionic) site occupations. Mott phases can occur for
commensurate filling (we take
$\av{n_{bi}}+\av{n_{fi}}=1$).


When the interactions are large enough, $U_{bf}\gg t_f,t_b$
and $U_{bb}\gg t_b$, fluctuations of the density are strongly
suppressed and the system is deep in the Mott phase. After
eliminating perturbatively the empty and doubly occupied
states one obtains a purely fermionic
model\cite{Lewenstein2004} \be H=-t_{\text{eff}}\sum_{\av{ij}} (c\yd_i
c\nd_j+\text{H.c.})+V_{\text{eff}}\sum_{\av{ij}} n_i n_j,
\label{Hf}
\ee
where $t_{\text{eff}}=t_f t_b/U_{bf}$ and
$V_{\text{eff}}=(t_b^2+t_f^2)/U_{bf}-2t_b^2/U_{bb}$. Here we have taken the singly occupied bosonic site as the
reference vacuum $\ket{\W}$.  The auxiliary fermion $c\yd_i=f\yd_i b_i$ is a quadratic operator in the original fields with  $c\yd_i\ket{\Omega}= f\yd_i b_i \ket{\Omega}$.
The ground state of the fermion  model~(\ref{Hf}) can be a Fermi
liquid for $V_{\text{eff}}\ge 0$, a $p$-wave superfluid for
small or moderatly large
negative $V_{\text{eff}}$, or be unstable to phase separation
 for large negative $V_{\text{eff}}$,
$|V_{\text{eff}}| \gg t_{\text{eff}}$.
 A phase transition out of the Mott state is driven by reducing the strength of the interactions $U_{bb}$, $U_{bf}$, or both.

 Consider first tuning the transition by changing $U_{bf}$, while
 $U_{bb}$ remains very large. For the resulting hard-core bosons it is
 useful to rewrite the problem in terms of holes in the Mott
 insulator, $h\yd_i=b_i$. The condition of unity filling reads now $\av{n_{hi}}=\av{n_{fi}}$ and the
repulsive interaction is mapped to attraction, $U_{bf}\to
 -U_{bf}$. In these variables, the Mott transition can be understood as binding of bosons
 to fermions, with the Mott state being a Fermi liquid of the
 molecules $c\yd_i\ket{\W}= f\yd_i h\yd_i\ket{\W}$  as $V_{\text{eff}}>0$ in
 this limit.

 A similar transition, from a Fermi surface of atoms to a Fermi
 surface of molecules, has been considered in
 Ref.~[\onlinecite{Powell2005}] for a Bose-Fermi mixture in the
 continuum with an interspecies Feshbach resonance. Our lattice model
 with $U_{bb}\gg U_{bf}$ maps to this continuum problem for low
 densities, $\av{n_{fi}}\ll 1$. We therefore expect two transitions
 as found in [\onlinecite{Powell2005}]. First a Fermi sea of molecules
 starts to form beyond a critical value of the attraction
 $U_{bf}=U_{c1}$ and coexists with the atomic Fermi sea and with a
 BEC of the unbound bosons. The volume of the molecular
 Fermi surface grows continuously until it reaches the full Luttinger
 volume, corresponding to the full fermion density, at
 $U_{bf}=U_{c2}$. At this point the condensate fraction vanishes and
 the system has reached the Mott insulating state. As pointed out in  Ref. [\onlinecite{Powell2005}], interactions are irrelevant at such a QCP in $d=3$ and due to the quadratic dispersion at the bottom of the bosonic and fermionic bands,
 $\omega \propto k^2$, the dynamical critical exponent is $z=2$.
The same theory can be applied for nearly unity filling by
 the fermions if we apply the particle hole transformation
 on the fermions rather than the bosons. Other phases with broken lattice symmetry are possible at certain intermediate fillings \cite{Buchler2003,Titvinize2008}.

We now turn to the main focus of this paper and consider the transition driven by reducing the boson-boson interaction $U_{bb}$ while keeping $U_{bf}$ large.
In this case we can eliminate the fermionic doublon state $f\yd_ib\yd_i\ket{0}$ at energy $\sim U_{bf}$. Using again the
 single boson state $\ket{\Omega}$ as the reference vacuum, we introduce besides
the single fermion state $c\yd_i\ket{\W}$ and bosonic hole $h\yd_i\ket{\Omega}$ defined above also the bosonic doublon  $p\yd_i\ket{\Omega}=2^{-\half}  b\yd_i\ket{\Omega}$.
The Hamiltonian (\ref{Hmic}) projected to low energies becomes
\bea \label{heffd}
H_{\rm eff}&=&\half U_{bb}\sum_i (n_{pi}+ n_{hi})- \mu_f\sum_i n_{c i} \label{Heff}\\
&&-t_b\sum_{\av{ij}} \left[(\sqrt{2}p\yd_i+h\nd_i)(\sqrt{2} p\nd_j + h\yd_i)+\text{H.c.}\right]\nn\\
&&-t_f\sum_{\av{ij}}
\left(c\yd_i h\nd_i h\yd_j c\nd_j+\text{H.c}\right)+U_{cc} \sum_{\av{ij}}
n_{ci} n_{cj} \nn
\eea
supplemented with the hard-core condition
$n_{pi}+n_{hi}+n_{ci}\le 1$ on each site.  $U_{cc}=(t_b^2+t_f^2)/U_{bf}$ describes an effective nearest-neighbor repulsion of the composite fermions.
Unity filling implies $\av{n_{hi}}=\av{n_{pi}}$ and the transition from the Bose-Fermi Mott state to the
superfluid is described as  a simultaneous condensation of the doublons and holes just as in a conventional bosonic Mott-superfluid transition \cite{Altman2002}.


{\em Critical theory --} To investigate the quantum critical point we analyze the corresponding critical theory.
The most general action allowed by symmetry (including only the most relevant terms) is given by
\bea
S&=& S_b+S_f+S_{bf}\nn\\
S_b&=&\int \! dx d\tau\, |\partial_\tau\phi|^2+c^2|\nabla\phi|^2+r|\phi|^2+u_b|\phi|^4\nn\\
S_f&=&\int  \bar{\psi}(\partial_\tau+ \vec{v}_F\cdot (- i \nabla-\vec k_F))\psi - u_f\bar\psi\nabla \bar\psi \cdot \psi\nabla \psi \nn\\
S_{bf}&=&u_{bf}\int \! dx d\tau \, \bar{\psi}\psi |\phi|^2
\label{Seff}
\eea
Here the bosonic order parameter field is related to the particle-hole fluctuations through $\phi(x) \sim \sqrt{1/c}(h(x)+ p\yd(x))$. As in the conventional bosonic Mott transition, the condition $\av{n_{hi}}=\av{n_{pi}}$ entails the absence of a linear time derivatives $\phi^* \partial_\tau\phi$. Formally the same theory was considered by Yang
\cite{Yang2008} to address the Mott transition at integer {\em boson} filling in contact with a fermi sea at a {\em non-commensurate} filling. The crucial difference here is that in our theory the $\psi$ fermions are not the physical atoms but rather composite objects consisting of a fermion atom and a bosonic hole. In the supplementary material we discuss how the parameters in Eq.~(\ref{Seff}) are related to the original microscopic model.

The bosonic sector of the field theory (\ref{Seff}) is identical to that of the commensurate Mott transition in the purely bosonic system characterized by a linear energy-momentum relation, $\omega \sim c k$, resulting in $z=1$.  We analyze the coupling of the bosonic to the fermionic sector by performing a scaling analysis using the decoupled fixed point $u_f=0$ and $u_{bf}=0$ as the reference point. Assuming  $t_f\ll t_b$, the couplings $u_f$ and $u_{bf}$ remain small. As the Fermi surface is two-dimensional while $\phi$ condenses only at $q=0$ there is no unique way to perform such a scaling analysis. Using in $d=3$ isotropic scaling, $\vec r \to \lambda \vec r$, $\tau \to \lambda \tau$, $\phi \to \phi/\lambda$ and $\psi \to \psi/b^{3/2}$,  suggests that $u_{bf}$ is irrelevant, $u_{bf} \to u_{bf}/\lambda$. Also alternative scaling schemes which take into account the curvature of the Fermi surface (following, e.g., Ref.  [\onlinecite{Metlitski2010}]) and direct calculation of leading diagrams show that $u_{bf}$ is irrelevant.
This does, however, not imply that it can be set to zero as $u_{bf}$ can generate a marginal long-ranged interaction of the bosons, see below. $u_f$ is marginal and leads in the attractive case, $u_f>0$,  to p-wave superfluidity.

We first assume, that the pairing instability of the Fermi surface can be neglected as either $u_f$ is repulsive or so small that the transition temperature is smaller than $T$. Due to the irrelevance
of $u_{bc}$ one can integrate out the fermions perturbatively to obtain a purely bosonic theory with a modified quartic interaction term
\be
S_{\rm int}=\! \int \!\! \left(u_0 + u_1 f( \w /{v_F q})\right)
\phi^*_{k-q,\nu-\w}\phi^*_{k'+q,\nu'+\w}\phi_{k',\nu'}\phi_{k,\nu}.
\ee
The $\omega$ and $q$ dependence of the new interaction vertex
\be
f(x)= {ix\over 2} \ln \left({i x+ 1\over i x -1}\right).
\ee
is inherited from the fermionic density-density correlation function.
We obtain
$u_1\approx  u_{bf}^2 \nu(0)=64 c^2 v_F  = {4\over \pi} {v_F\over c}~ u_b$
where $\nu(0)$ is the fermion density of states. The
local interaction $u_0$ also receives a correction, $u_0\approx u_b +u_1$.

To investigate the fate of the critical point we set up a perturbative renormalization group (RG) by integrating out momenta with $\Lambda/b<|q|<\Lambda$. A rescaling,  $k\to k/b$ and $\w\to \w/b^z$, restores the original cutoff $\Lambda$.  Due to the $\omega$ dependence of the interactions, already to one-loop order one obtains self energy corrections which are absorbed by rescaling of the field $\phi$ and a correction to the dynamical critical exponent $z$  to keep the quadratic part of the action fixed.
A complication of the scheme is that
higher-order long range terms of the form $u_n f(\w/ v_f q)^n$ are generated during the RG flow. The scaling is therefore determined by coupled equations for the dimensionless coupling constants $g_n(b)=u_n(b)/(8\pi^2 c^3)$ and the dimensionless Fermi velocity $\eta(b)=v_F(b)/c$.
\bea
\frac{d \eta}{d l}&=&\frac 2 3 \sum_{m=1}^\infty  f^\eta_m(\eta)\, g_m\\
\frac{d g_0}{d l}&=&-10 \,g_0^2-12 g_0 \sum_m
f^g_{m}(\eta)\, g_m -4\sum_{m,n=1}^{\infty}f^g_{m+n}(\eta) g_m g_n  \nonumber \\
\frac{d g_n}{d l}&=& -2 \sum_{m=0}^n  g_{n-m}\,  g_m  - 4 g_n \sum_{m=0}^\infty
f^g_{m}(\eta)\, g_m, \quad {\rm n>0},\nonumber
\label{eq:RG}
\eea
where $f^\eta_m(\eta)=\frac{4 \eta}{\pi} \int_0^\infty \frac{3 \eta^2 x^2-1}{(\eta^2 x^2+1)^3} f(x)^m d x$ and $f^g_m(\eta)=\frac{4 \eta}{\pi} \int_0^\infty \frac{1}{(\eta^2 x^2+1)^2} f(x)^m d x$ are functions of $\eta(b)$, $l=\ln b$.

\begin{figure}[t]
\centerline{\resizebox{1.0\linewidth}{!}{\includegraphics{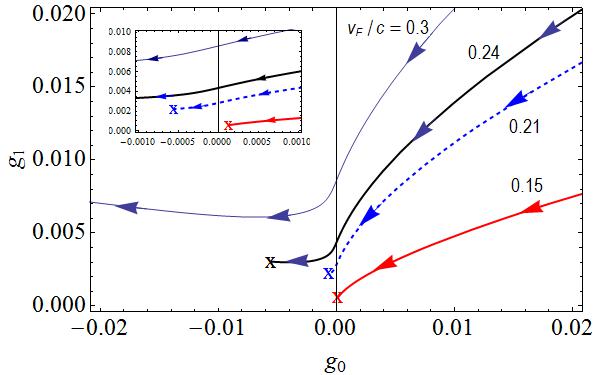}}}
 \caption{RG flow for three different values of the bare Fermi velocity $v_F/c$ using $n_{max}=13$ (see text). The inset is a zoom near $g_0=0$ showing that for the three larger values of $v_F/c$ the local interaction is driven to negative values before the flow is cut off at the scale of the pairing gap $\D=\e_F \exp(-8\pi c/v_F)$. For these values we expect a fluctuation driven first order transition.}
 \label{fig:RG}
\end{figure}

In solving for the flow we keep terms with $n<n_{\text{max}}$ and find that the resulting flow converges with $n_{\text{max}}$, keeping about 10 terms is enough in practice. Typical RG trajectories are shown in Fig.~\ref{fig:RG}.
While higher couplings $g_n$ and the $\eta$ dependence is quantitatively important,the qualitative structure of the flow can be understood by considering only $g_0$ and $g_1$ for fixed $\eta$.
Initially, both $g_0$ and $g_1$ drop but the flow of $g_1$ is much slower due to its non-local nature. Therefore $g_1$ is finite when $g_0$ reaches zero, driving $g_0$ to negative values through the last term in the flow equation (\ref{eq:RG})for $g_0$, thus leading to a first order transition.

In the RG approach described above we have neglected the induced attraction $u_f$ between fermions, which would lead to a pairing instability and opening of a gap $\D\approx E_F \exp(-8\pi c/v_F)$ in the Fermi surface. Such a gap will suppress the non-local coupling terms at low energies $c\L/b<\D$. Therefore, if the coupling constants $g_n$ have not yet driven the local coupling $g_0$ to negative values at that scale, the first order transition will be avoided.
Numerical solution of the flow equations suggest that this is the case if the bare ratio $v_F/c< 0.18$, while for $v_F/c>0.18$ we expect a first order transition.
In either case the Fermi surface is expected to give way to a small $p$-wave gap near the transition for $T \to 0$.

Our analysis applies to a number of other quantum critical points where bosonic and fermionic degrees of freedom coexist. Consider, for example, a metallic commensurate antiferromagnet, where the shape of the Fermi surface is such, that the ordering wave vector $\vec Q$ does not connect different parts of the Fermi surface. In such a situation, there is no linear coupling, $\phi \psi\yd \psi$, of the order parameter $\phi$ to low-energy fermions and only a quadratic coupling of fermions to a $z=1$ bosonic QCP survives. Our analysis shows that in $d=3$ this coupling will render the quantum critical point always weakly first order as long as no superconductivity gaps out the Fermi surface.

{\em Pseudogap --} We now discuss the experimental ramifications of the quantum phase transition focussing on the spectral function associated with emission of a fermionic atom in RF spectroscopy \cite{Jin2008,Kohl2011}.
The crucial point to note is that in the long-wave length limit the physical atomic fermions $f_i$ are composite objects  in terms of the weakly coupled fields $\psi$
and $\phi$, $f(x)\sim \sqrt{1/c} \phi(x)\psi(x)$ as $f_i=h\yd_i c_i$. Hence, the spectral function should be found from the Green's function $\mathcal{G}(x,\tau)=\av{\bar\psi(x,\tau)\phi^*(x,\tau)\phi(0) \psi(0)}$.

For the sake of this discussion we ignore all logarithmic corrections
which ultimately lead either to  p-wave pairing or the fluctuation induced first order transition. These subtle effects are only noticeable at exponentially energies.
The salient features of the spectral function at higher energies (or temperatures) are captured within the Gaussian theory obtained from expansion about the saddle point of (\ref{Seff}), which implies $\mathcal{G}(x,\tau)=G_\psi(x,\tau)\mathcal{D}_{\phi}(x,\tau)$, where $G_\psi$ is the free fermion Green's function associated with the composite fermions.

\begin{figure}[t]
 \centering
 \begin{tabular}{cc}
 \includegraphics[width=0.99 \linewidth]{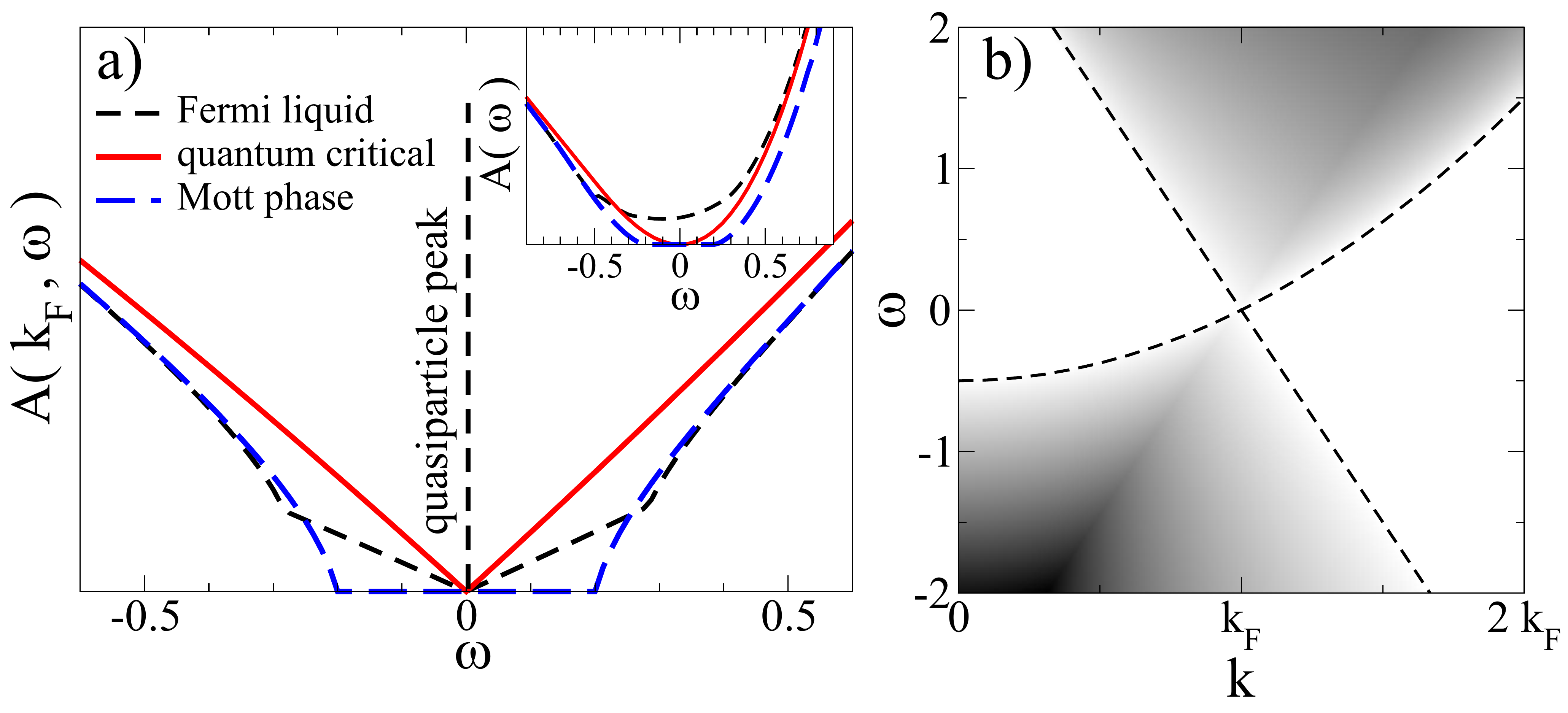} & 
 \end{tabular}
 \centering
  \caption{(a) Spectral function $A(k,\w)$ of the fermions  at $k=k_F$ in the superfluid phase, critical point and Mott insulator ($r=-0.04,0,0.04$, $c/v_F=3$). Inset:  local density of states $A(\w)=\int dk A(k,\w)$ in the three cases. (b) The spectral function $A(k,\w)$ at the critical point.}
 \label{fig:Akw}
\end{figure}

In the superfluid side, we can take a Bogoliubov expansion of the order parameter $\phi=\phi_0+\d\phi_1+ i\d\phi_2$, to split the bosonic component of the Green's function into three contributions:
$\mathcal{D}_\phi(x,\tau)=|\phi_0|^2+\mathcal{D}_1(x,\tau)+\mathcal{D}_2(x,\tau)$. The condensate part $|\phi_0|^2$, when combined with the free fermion Green's function $G_\psi$, will give a delta function contribution of magnitude $|\phi_0|^2$ dispersing with the free fermion dispersion.
The phonon contribution leads to a continuous spectrum  rising linearly with $\w$ up from the energy of the quasi-particle peak.
Another continuous contribution onsets above the energy gap of the amplitude (or Higgs) mode. All three features are seen Fig. \ref{fig:Akw}(a), where the
spectral function $A(k,\w)$ at $k=k_F$ has been calculated for $c/v_F=3$ and
a quadratic fermionic dispersion $\epsilon_k=k^2/2$.

Upon approaching the
critical point, the quasiparticle weight $Z \sim |\phi_0|^2 \sim U_{bb}^c-U_{bb}$
and the associated discontinuity in the momentum distribution function of the fermions $\av{n^f_\bk}$
decrease to zero. Correspondingly, a pseudogap develops in the local density of states, see inset of Fig.~\ref{fig:Akw}(a). Directly at the critical point, $U_{bb}=U_{bb}^c$, where $Z=0$, the spectral function at $k=k_F$ rises linearly in $\w$, see Fig.~\ref{fig:Akw}(a).
The underlying quadratic dispersion of the composite fermions and the linear dispersion of the bosonic excitations are clearly visible in  Fig. \ref{fig:Akw}(b), which shows $A(k,\w)$ at criticality, with $k$ varying along a radial direction.

Finally,
inside the Mott phase the bosonic fluctuations can be treated as a free massive field. Hence upon convolution with the fermion Green's function one obtains a fully gapped spectral function despite the existence of a gapless Fermi liquid.

The gapless fermions that exist in the Mott insulator are hidden from standard single particle probes such as photoemission or the momentum distribution measured in time of flight. Interestingly however, the hidden Fermi surface can be revealed by noise correlations in time of flight images\cite{Altman2004a}. The boson-fermion cross correlations at momenta $\bk$ and $\bk+\bq$ are directly proportional to the momentum-$\bq$ distribution of the  {\em composite fermions},
\be  \av{n^c_{\bq}} \approx \sum_{\bk} \av{n^f_{\bk+\bq} n^b_{\bk}}-\av{n^f_{\bk+\bq}}\av{n^b_{\bk}}. \ee
The approximation becomes exact deep in the Mott insulating state, where $f^\dagger_i b_j = \delta_{ij} c_i^\dagger$ for $U_{bf}, U_{bb}\to \infty$.

The composite fermions can also be seen through Bragg or lattice modulation probes that couple with different strength to the bosons and fermions. In this case they effectively couple to the composite fermions. The appropriate structure factors will therefore display a gapless spectrum in the composite Fermi liquid phase. Onset of p-wave pairing of the composite fermions, expected in certain regions of the Mott phase, will be seen as opening of a gap with a singular spectrum at the gap edge.

{\em Conclusions --}
Mixed boson-fermion systems in optical lattices open a new route, for both theoretical and experimental investigation of unconventional Mott transitions that entail the destruction of
fermionic quasi-particles and the emergence of hidden Fermi surfaces of composite particles.
Our theory accounts for the critical behavior of the single fermion spectral function and gives a simple and tractable example for the emergence of a pseudogap in a strongly correlated system.

 At very low energy scales, coupling of the critical modes to the Fermi surface can either lead to p-wave pairing of the composite Fermi surface or can destabilize the critical point leading to a fluctuation induced first order transition. While for $d=3$, the nature of the phase transition depends
on the ratio the interactions, see  Fig. \ref{fig:schem}, it has been shown that in $d=1$ always a single $z=1$ transition is
expected\cite{Mathey2012}.

An interesting open question is the nature of the possible tri-critical point postulated in Fig.  \ref{fig:schem} where $U_{bb}\approx U_{bf}$ and the two $z=2$ transitions meet with the $z=1$ critical point. Interestingly, the generalized Hubbard model~(\ref{Hmic}) exhibits supersymmetry at  $U_{bb}=U_{bf}$, $\mu_b=\mu_f$ and $t_f=t_b$~\cite{Yu2008}. The ground state at the super-symmetric point has no fermions but may help to elucidate the nature of the tri-critical point in the limit of low fermion density. Another open question concerns the Mott transition in a commensurate mixture of bosons and spinfull fermions.

We acknowledge useful discussion with S. Trebst, E. Berg, D. Podolsky and financial support by the DFG (FOR 960, SFB 608), ISF under grant 1594/11 (E.A.), and the U.S. Israel BSF (E.A. and E.D.), DARPA OLE program, Harvard-MIT CUA, NSF Grant No. DMR-07-05472,
AFOSR Quantum Simulation MURI, the ARO-MURI on Atomtronics (E.D.).

\newpage {\bf Supplementary material:}

\section{Parameters of the effective field theory}

Derivation of the field theory from the lattice model~(\ref{Heff}) involves integration over the short wavelength modes up to an intermediate scale $\Lambda_{GL}$, much smaller than the bare cutoff $\Lambda_0$ (set by the inverse lattice spacing $a$), but much larger than the inverse correlation length of the QCP.
In general, it is difficult to obtain precise values of the parameters of the effective low-energy theory, but at least for small fermion density, $\av{n_{pi}}\ll 1$, it is simple to estimate their order of magnitude. $r \sim U_{bb}-U_{bb}^c$ is determined by the distance to critical interaction strength $U_{bb}^c$ of the order of the bosonic bandwidth, which also determines the velocity of the critical mode, $c \sim t_b$.
The hard core interaction between the particles at scale $\Lambda_0$ is renormalized to a finite effective repulsion, $u_b \approx 8\pi^2 c^3$, given by the scattering $T$-matrices between bosons at the scale $\Lambda_{GL}$  (we have taken $\ln(\L_0/\L_{GL})\sim 1$). The kinetic energy of the fermions is generated from the correlated hopping term proportional to $t_f$ in (\ref{Heff}) following integration over the high energy boson holes and is proportional to the hole density in the lattice which is a finite number of order $1$ at the QCP. In other words the effective mass of the Fermions is $1/m_* \sim  t_f \av{n_h}\sim t_f$. Using the Free fermion propagator thus generated, we can now find the magnitude of the boson-fermion vertex in the same way as we did for the bose-bose vertex and obtain $u_{bf}\approx 8\pi c/m_*$. Finally, the integration over the holes also generates an effective p-wave attraction between the fermions, $u_{f}\sim 1/(8\pi m_*^2 c)$ for $t_f \ll t_b$. The dimensionless interaction strength is then $u_{f} \nu(E_F)\sim v_F/(8\pi c)$ for  $t_f \ll t_b$ and of order $1$ for  $t_f \gtrsim t_b$.
For $U_{bf} \gg U_{bb}$ this attractive interaction will be larger than the small repulsive interaction arising from $U_{cc}$ in Eq.~(\ref{Heff}).

\end{document}